\documentclass[preprint,superscriptaddress,amsmath,amssymb,prd,aps,showpacs,floatfix,nofootinbib]{revtex4-1}
\usepackage{graphicx}
\usepackage{dcolumn}
\usepackage{bm}
\usepackage{mathrsfs}
\usepackage{hyperref}
\usepackage{amsmath}
\usepackage{amssymb}
\usepackage{bm}
\usepackage{color}
\usepackage{float}
\usepackage{dcolumn}
\usepackage{multirow}
\usepackage{changepage}
\usepackage{enumerate}
\usepackage{ulem}
\usepackage[dvipsnames]{xcolor}
\usepackage{braket}
\usepackage{nicefrac}
\usepackage{multirow}
\usepackage{tabularx}
\usepackage{subfigure}
\usepackage{soul}
\usepackage[makeroom]{cancel}

\hypersetup{pdftex,colorlinks=true,linkcolor=blue,citecolor=ForestGreen,menucolor=black,urlcolor=blue,filecolor=blue}

\newcommand{\mev}{\textrm{ MeV}}

\begin{document}

\title{$\bar{B}_s^0 \to D_{s1}(2460)^+ K^-, D_{s1}(2536)^+ K^-$ and the nature of the two $D_{s1}$ resonances}
\date{\today}

\author{Jia-Xin Lin}
\affiliation{School of Physics, Southeast University, Nanjing 210094, China}

\author{Hua-Xing Chen}
\email{hxchen@seu.edu.cn}
\affiliation{School of Physics, Southeast University, Nanjing 210094, China}

\author{Wei-Hong~Liang}
\email{liangwh@gxnu.edu.cn}
\affiliation{Department of Physics, Guangxi Normal University, Guilin 541004, China}
\affiliation{Guangxi Key Laboratory of Nuclear Physics and Technology, Guangxi Normal University, Guilin 541004, China}

\author{Chu-Wen Xiao}
\email{xiaochw@gxnu.edu.cn}
\affiliation{Department of Physics, Guangxi Normal University, Guilin 541004, China}
\affiliation{Guangxi Key Laboratory of Nuclear Physics and Technology, Guangxi Normal University, Guilin 541004, China}

\author{Eulogio Oset}
\email{oset@ific.uv.es}
\affiliation{Department of Physics, Guangxi Normal University, Guilin 541004, China}
\affiliation{Departamento de F\'{\i}sica Te\'orica and IFIC, Centro Mixto Universidad de
Valencia-CSIC Institutos de Investigaci\'on de Paterna, Aptdo.22085,
46071 Valencia, Spain}

\begin{abstract}
    Starting from the molecular picture for the $D_{s1}(2460)$ and $D_{s1}(2536)$ resonances, which are dynamically generated by the interaction of coupled channels, the most important of which are the $D^*K$ for the $D_{s1}(2460)$ and $DK^*$ for the $D_{s1}(2536)$, we evaluate the ratio of decay widths for the $\bar{B}_s^0 \to D_{s1}(2460)^+ K^-$ and $\bar{B}_s^0 \to D_{s1}(2536)^+ K^-$ decays, the latter of which has been recently investigated by the LHCb collaboration, and we obtain a ratio of the order of unity. The present results should provide an incentive for the related decay into the $D_{s1}(2460)$ resonance to be performed, which would provide valuable information on the nature of these two resonances.
\end{abstract}

\maketitle

\section{Introduction}
\label{sec:intro}
The $D_{s1}(2460)$, $D_{s1}(2536)$ are axial vector resonances, $J^P = 1^+$, containing a $c$ quark and a strange antiquark.
The axial vector resonances have attracted the attention of the hadron community since many of them can be interpreted as molecular states coming from the interaction of vector mesons with pseudoscalars \cite{Lutz:2003fm,Roca:2005nm,Geng:2006yb,Garcia-Recio:2010enl,Zhou:2014ila}.
This picture becomes more attractive from the perspective of quark models being less accurate in this than in other sectors of the meson spectrum \cite{Godfrey:1985xj}.
In the charm sector, the lightest axial vector resonances are the $D_{s1}(2460)$ and $D_{s1}(2536)$ reported by many experimental groups \cite{PDG:2022ynf}.
The $D_{s1}(2460)$, in connection with its spin partner $D_{s0}^*(2317)$, has been the subject of intense debate. The simplest thing is to assume that these states are ordinary $q \bar q$ states, 
which has been advocated in Refs.~\cite{Bardeen:2003kt,Nowak:2003ra, Dai:2003yg, Lucha:2003gs, Sadzikowski:2003jy, Becirevic:2004uv, Lee:2004gt, Wei:2005ag, Liu:2013maa, Wang:2006mf, Colangelo:2005hv}.
However, the four quark structure has also been suggested in many works \cite{Cheng:2003kg,Terasaki:2003qa, Browder:2003fk, Dmitrasinovic:2005gc, Hayashigaki:2004gq, Bracco:2005kt, Kim:2005gt, Nielsen:2005ia,Esposito:2014rxa}. 
One reason to advocate for more complex structures than the ordinary $q\bar q$ is the failure of the, otherwise successful, relativised quark model \cite{Godfrey:1985xj}, 
which predicts higher masses than the experimental ones.
On the other hand, it was soon realized that the $DK, D^*K$ channels can play an important role in lowering the mass of these states \cite{vanBeveren:2003kd, Dai:2003yg, Guo:2006fu, Lang:2014yfa}. 
The realization of this point, together with the success of the chiral unitary approach, describing scalar and axial vector mesons in the SU(3) sector as molecular states stemming from the meson meson interaction \cite{Oller:2000ma}, 
gave rise to the molecular interpretation of these states, 
which has met with a large support \cite{Guo:2006fu, Xie:2010zza,Zhang:2006ix,Bicudo:2004dx,Kolomeitsev:2003ac,Liu:2012zya,Cleven:2014oka, Gamermann:2006nm, Gamermann:2007fi, Altenbuchinger:2013vwa,MartinezTorres:2018zbl, Guo:2006fu,Guo:2006rp,Barnes:2003dj,Cahn:2003cw, Altenbuchinger:2013vwa,Yao:2015qia,Guo:2015dha, Du:2017ttu,Gil-Dominguez:2023huq,Wu:2022wgn}.
This molecular picture, in which the $DK$ and $D^*K$ channels are dominant in the interpretation of the $D_{s0}^*(2317)$ and $D_{s1}(2460)$ states, 
has been widely used studying strong and radiative decays of these resonances in Refs.~\cite{Faessler:2007gv, Faessler:2007us, Cleven:2014oka, Xiao:2016hoa, Lu:2006ry, Wei:2005ag}. 
It has also been tested in production experiments from decays of $B$ states \cite{Datta:2003re, Liu:2022dmm, Navarra:2015iea, Albaladejo:2016hae}. 
Some other works advocate a mixture of $q \bar q$ and molecular components \cite{Albaladejo:2018mhb}.
Lattice QCD calculations have also come to give support to the molecular picture \cite{Liu:2012zya, Lang:2014yfa, Cheung:2020mql},
although some of them also advocate for a mixture of $q \bar q$ and molecular components \cite{Yang:2021tvc}.

As we can see, there are many different interpretations on the nature of these resonances, although the molecular picture has a stronger support, 
and, whatever original picture one uses, the appearance of molecular components seems unavoidable. This has been made explicit in Refs.~\cite{Dai:2023kwv,Song:2023pdq}, 
showing that even starting from a nonmolecular state that generates a bound state close to a threshold of two particles, the state become purely molecular in the limit of zero binding. 
For small but finite binding energy, forcing the state to have a small molecular component reverts in abnormally large effective range parameters and very small scattering lengths, that can be ruled out by experiment.

With this panorama, any new idea that can shed light on this issue is most welcome. 
From this perspective we propose here a reaction that can provide new information on the nature of the two axial vector resonances, $D_{s1}(2460)$ and $D_{s1}(2625)$.

In the molecular picture \cite{Kolomeitsev:2003ac,Gamermann:2007fi}, the $D_{s1}(2460)$ and $D_{s1}(2536)$ are obtained from the interaction of coupled channels, but the relevant ones are $D^* K$ for the $D_{s1}(2460)$ and $DK^*$ for the $D_{s1}(2536)$.
One may be surprised that the binding energy is of the order of $40\mev$ for the $D^* K$ state, while it is of the order of $200\mev$ for the $DK^*$ one.
In the extension of the local hidden gauge approach \cite{Bando:1984ej,Bando:1987br,Meissner:1987ge,Nagahiro:2008cv} to the charm sector \cite{Montana:2017kjw,Debastiani:2017ewu}, using the exchange of vector mesons as a source of the interaction, this has a natural interpretation, since close to threshold the interaction is proportional to $m_K$ for the $D^* K$ interaction, while it is proportional to $m_{K^*}$ for the $DK^*$ interaction and this factor difference leads, indeed, to a much bigger binding of the $DK^*$ channel.

Lattice QCD calculations have also brought light into this issue, and the lattice data of Ref.~\cite{Lang:2014yfa} are analyzed in Ref.~\cite{MartinezTorres:2014kpc} and conclude that the $D_{s1}(2460)$ state is mostly of $D^* K$ molecular nature with a probability of this component of $(57 \pm 21 \pm 6)\%$.
Similar results are obtained starting from a quark model looking for the overlap with molecular components in Ref.~\cite{Ortega:2016aao}. 

The purpose of the present work is to dig further into this issue and propose one new test for the molecular picture.
We take advantage of the recent experimental measurement by the LHCb collaboration of the $B_s^0 \to D_{s1}(2536)^\mp K^\pm$ decay \cite{LHCb:2023eig} and we make predictions for the decay $B_s^0 \to D_{s1}(2460)^- K^+$.
The rates of these two decays can be accurately calculated within the molecular picture hence providing a strong test of the molecular nature of these resonances.

\section{Formalism}
\label{sec:form}
\begin{figure}[htb]
    \centering
    \subfigure[]{\includegraphics[scale=0.52]{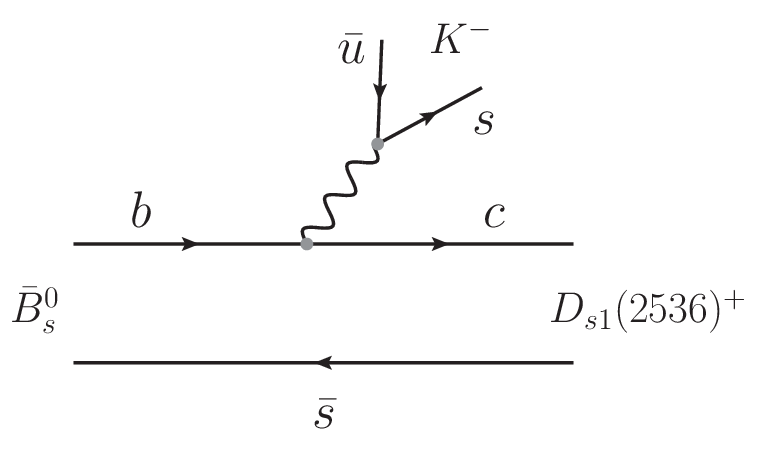}\label{fig:Bdecay-a}}~~~~~
    \subfigure[]{\includegraphics[scale=0.52]{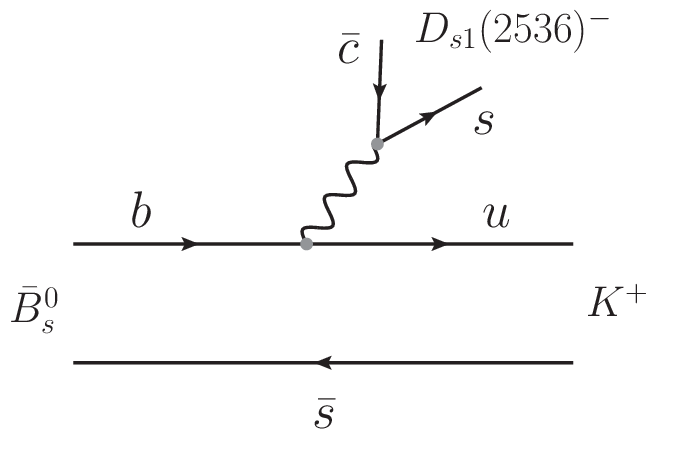}\label{fig:Bdecay-b}}
    \caption{Mechanisms for $\bar{B}_s^0 \to D_{s1}(2536)^\pm K^\mp$ in Ref.~\cite{LHCb:2023eig}.}
    \label{fig:Bdecay}
\end{figure}

In Ref.~\cite{LHCb:2023eig} the $B_s^0 \to D_{s1}(2536)^\mp K^\pm$ decay was measured and the mechanisms of Fig.~\ref{fig:Bdecay} were proposed, with Cabibbo suppressed mechanisms (we plot the diagrams with $\bar{B}^0_s$ to deal with $b$ quarks).

By looking at the Kobayashi-Maskawa matrix elements \cite{Kobayashi:1973fv} we see that the Fig.~\ref{fig:Bdecay-a} requires $V_{bc} V_{us} = 0.0092$ and Fig.~\ref{fig:Bdecay-b} goes with $V_{ub} V_{cs} = 0.0037$.
We will then choose the most favored process of Fig.~\ref{fig:Bdecay-a} to study the relationship of $D_{s1}(2460)$ and $D_{s1}(2536)$ production, hence we will study the $\bar{B}_s^0 \to D_{s1}(2460)^+ K^-$ and $\bar{B}_s^0 \to D_{s1}(2536)^+ K^-$ reactions.

From the molecular perspective the $D_{s1}(2460)$ and $D_{s1}(2536)$ resonances are not a $c\bar{s}$ state, as implicitly assumed in Fig.~\ref{fig:Bdecay-a}, but they stem from the vector meson-pseudoscalar interaction of coupled channels. 
In Ref.~\cite{Gamermann:2007fi} several channels were considered that we reproduce in Table~\ref{tab:axial}, together with the obtained couplings of the state to the different channels.
\begin{table}[b]
    \centering
    \renewcommand{\arraystretch}{1}
    \caption{The results for the poles in the $C=1, S=1, I=0$ sector and the couplings of the two poles to different channels, taken from Ref.~\cite{Gamermann:2007fi}.}
    \begin{tabular}{c|c|c}
    \hline\hline
    \,\,\multirow{2}*{channel}\,\, & \,$2455.91 \mev$\, & \,$(2573.62-i 0.07) \mev$\,
    \\ 
    & $|g_i|(\rm GeV)$ & $|g_i|(\rm GeV)$
    \\\hline\hline
    $D K^*$ & 0.54 & 13.96
    \\ \hline
    $K D^*$ & 9.74 & 0.30
    \\ \hline
    $\eta D_s^*$ & 6.00 & 0.18
    \\ \hline
    $D_s \omega$ & 0.51 & 7.95 
    \\ \hline
    $\eta_c D_s^*$ & 0.02 & 0.05
    \\ \hline
    $D_s J / \psi$ & 0.54 & 0.00
    \\ \hline\hline
    \end{tabular}
    \label{tab:axial}
\end{table}
As one can see, the relevant channels, according to the size of the couplings, are $D^* K$ and $D_s^* \eta$ for the $D_{s1}(2460)$ and $DK^*$, $D_s \omega$ for the $D_{s1}(2536)$.
We will neglect all the other channels and discuss the relevance of those that we keep.

The mechanism to produce the molecular states starts from the diagram of Fig.~\ref{fig:Bdecay-a} at the quark level, but the $c\bar{s}$ component has to be hadronized to give two mesons and allow the molecule to be formed through the final state interaction of these mesons.
Technically we proceed as follows: we introduce the $q\bar{q}$ matrices $P$, $V$ written in terms of the physical pseudoscalar and vector mesons, respectively, 
\setlength{\arraycolsep}{5pt}
\renewcommand{\arraystretch}{1.2}
\begin{eqnarray}
    P &=& \left(\begin{array}{cccc}
    \frac{\eta}{\sqrt{3}}+\frac{\eta^{\prime}}{\sqrt{6}}+\frac{\pi^0}{\sqrt{2}} & \pi^{+} & K^{+} & \bar{D}^0 \\
    \pi^{-} & \frac{\eta}{\sqrt{3}}+\frac{\eta^{\prime}}{\sqrt{6}}-\frac{\pi^0}{\sqrt{2}} & K^0 & D^{-} \\
    K^{-} & \bar{K}^0 & -\frac{\eta}{\sqrt{3}}+\sqrt{\frac{2}{3}} \eta^{\prime} & D_s^{-} \\
    D^0 & D^{+} & D_s^{+} & \eta_c
    \end{array}\right), \\[2mm]
    V &=& \left(\begin{array}{ccccc}
    \frac{\omega}{\sqrt{2}}+\frac{\rho^0}{\sqrt{2}} & \rho^{+} & K^{*+} & \bar{D}^{* 0} \\
    \rho^{-} & \frac{\omega}{\sqrt{2}}-\frac{\rho^0}{\sqrt{2}} & K^{* 0} & D^{*-} \\
    K^{*-} & \bar{K}^{* 0} & \phi & D_s^{*-} \\
    D^{* 0} & D^{*+} & D_s^{*+} & J / \psi 
    \end{array}\right),
\end{eqnarray}
with the $\eta-\eta^\prime$ mixing of Ref.~\cite{Bramon:1992kr}. 

\begin{figure}[b]
    \includegraphics[width=0.4\linewidth]{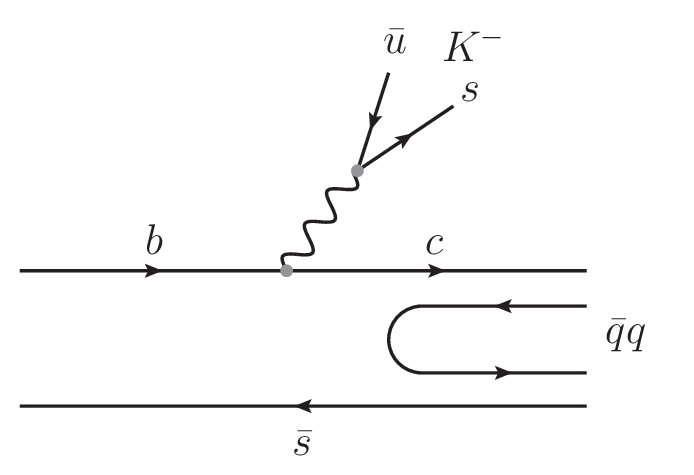}
    \vspace{-0.7cm}
    \caption{Mechanism of hadronization to produce $K^-$ and two extra mesons.}
    \label{fig:hadronic}
\end{figure}

Then the hadronization proceeds as shown in Fig.~\ref{fig:hadronic}
and, thus, 
\begin{equation}
    \label{eq:HPV}
    c\bar{s} \to H_1 = \sum\limits_{i} c \bar{q}_i q_i \bar{s} = \sum\limits_{i} P_{4i} V_{i3} =(PV)_{43},
\end{equation}
if we choose the $PV$ combination and
\begin{equation}
    \label{eq:HVP}
    c\bar{s} \to H_2 = \sum\limits_{i} c \bar{q}_i q_i \bar{s} = \sum\limits_{i} V_{4i} P_{i3} =(VP)_{43},
\end{equation}
if we choose the $VP$ one.
Then we have
\begin{eqnarray}
    \label{eq:PV}
    (PV)_{43} &=& D^0 K^{*+} + D^+ K^{*0} +D_s^+ \phi +\eta_c D_s^{*+}, \\[2mm]
    \label{eq:VP}
    (VP)_{43} &=& D^{*0} K^{+} + D^{*+} K^{0} +D_s^{*+} (\frac{-\eta}{\sqrt{3}}+\sqrt{\frac{2}{3}}\eta^\prime) +J/\psi D_s^{+}.
\end{eqnarray}
The $D_s \phi$ does not appear in Table~\ref{tab:axial}, the $\eta_c D_s^*$ has a negligible coupling in Table~\ref{tab:axial} and so is the case of the $J/\psi D_s^+$ channel.
The $\eta^\prime$ term is also inoperative due to its large mass and one has to discuss the relevance of the $D^{*+} \eta$ channel.
In Table~\ref{tab:axial} the relative weight to the $\frac{1}{\sqrt{2}}(D^{*+}K^0 + D^{*0}K^+)$ is about $6/10$, so with respect to the combination of Eq.~\eqref{eq:VP} we have a reduction factor of $\frac{6}{10} \cdot \frac{1}{\sqrt{3}} \cdot \frac{1}{\sqrt{2}}$, which together with the mass being about $200\mev$ higher than that of the $D_{s1}(2460)$, make again this channel negligible.
Hence, we have a clear case where we have a clean $D^* K$ or $DK^*$ production in $I=0$, where the final state interaction gives rise to the resonances.
The hadronizations in Eqs.~\eqref{eq:PV} and \eqref{eq:VP} indicate that $(PV)_{43}$, $(VP)_{43}$ are produced with the same weight.

\begin{figure}[b]
    \centering
    \subfigure[]{\includegraphics[width=0.4\hsize, height=0.2\hsize]{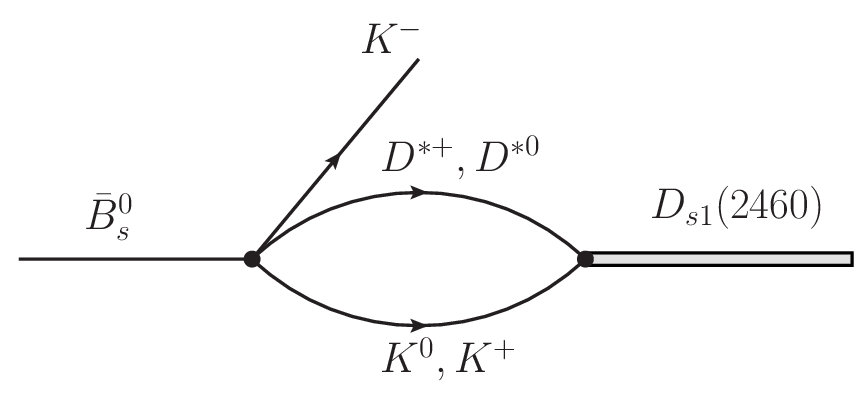}\label{fig:Ds12460}}
    \subfigure[]{\includegraphics[width=0.4\hsize, height=0.2\hsize]{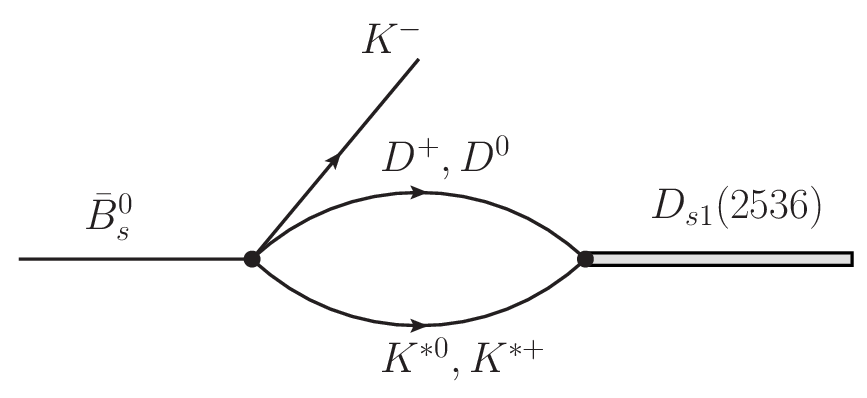}\label{fig:Ds12536}}
    \caption{Mechanism of final state interaction to produce the $D_{s1}(2460)$ and $D_{s1}(2536)$ resonances.}
    \label{fig:final}
\end{figure}

With the isospin multiplets convention $(D^+, -D^0)$, $(K^+, K^0)$, $(D^{*+}, -D^{*0})$, $(K^{*+}, K^{*0})$, the isospin states are given by
\begin{eqnarray}
    \label{eq:IKDs}
    \ket{D^* K, I=0} &=& \frac{1}{\sqrt{2}} (D^{*+} K^0 + D^{*0} K^+),\\[2mm]
    \label{eq:IDKs}
    \ket{D K^*, I=0} &=& \frac{1}{\sqrt{2}} (D^{+} K^{*0} + D^{0} K^{*+}).
\end{eqnarray}
Then, the production of the resonances proceeds as shown in Fig.~\ref{fig:final}.

Analytically we have
\begin{eqnarray}
    \label{eq:t2460}
    t_{\bar{B}_s^0 \to K^- D_{s1}(2460)^+} &=& \mathcal{C}^\prime \;G_{D^* K}(\sqrt{s}=2460) \;g_{D_{s1},D^*K}, \\[2mm]
    t_{\bar{B}_s^0 \to K^- D_{s1}(2536)^+} &=& \mathcal{C}^\prime \;G_{D K^*}(\sqrt{s}=2536) \;g_{D_{s1}^\prime,DK^*},
    \label{eq:t2536}
\end{eqnarray}
where $\mathcal{C}^\prime$ is a normalization constant, common to both decays, that will cancel in the evaluation of ratios,
$g_{D_{s1}, D^* K}$, $g_{D_{s1}^\prime,DK^*}$ are the couplings of the $D_{s1}(2460)$ and $D_{s1}(2536)$ to $D^* K$ and $D K^*$, respectively, and $G_{D^* K}(\sqrt{s})$ and $G_{D K^*}(\sqrt{s})$ are the loop function for intermediate $D^* K$ or $D K^*$ propagation which one has used in the derivation of the scattering matrix in coupled channels \cite{Gamermann:2007fi} 
\begin{equation}
    \label{eq:BS}
    T = [1-VG]^{-1}\,V,
\end{equation}
where $V$ is the transition potential evaluated via vector exchange using the extension of the local hidden gauge approach \cite{Bando:1984ej,Bando:1987br,Meissner:1987ge,Nagahiro:2008cv}.

\begin{figure}[b]
    \includegraphics[width=0.25\linewidth]{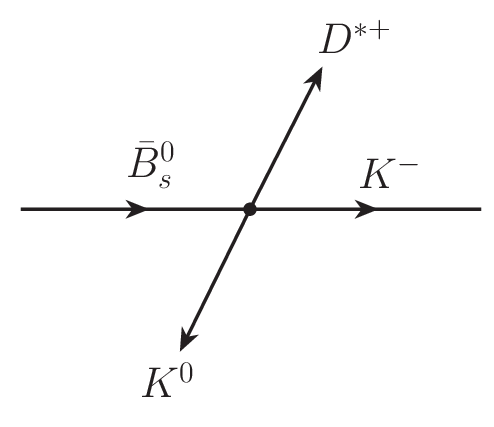}
    \vspace{-0.4cm}
    \caption{Momenta of the particles in the $D_{s1}(2460)$ rest frame.}
    \label{fig:momenta}
\end{figure}

There is one technical detail: in the rest frame of the $D_{s1}(2460)$ resonance the coupling to $VP$ goes as
\begin{equation}
    g_{D_{s1}} \, \vec{\epsilon}_{D_{s1}} \cdot \vec{\epsilon}_V.
\end{equation}
In this frame the $\bar{B}_s^0$ and $K^-$ have the same momentum and $D^{*+}$, $K^0$ have opposite momenta (see Fig.~\ref{fig:momenta}).
In the $\bar{B}_s^0 \to D^{*+} K^0 K^-$ vertex we must contract the $\vec{\epsilon}_{D^*}$ with a vector, which we take $\vec{p}_{\bar{B}_s^0} = -\vec{p}_K$. 

In Fig.~\ref{fig:Ds12460} we have to sum over polarizations of the vector $D^*$, and, thus, we have in the $D_{s1}(2460)$ rest frame
\begin{eqnarray}
    \sum_{\mathrm{pol}} p_{\bar{B}_s^0}^i \, \epsilon_{D^*}^i \, \epsilon_{D^*}^j \, \epsilon_{D_{s1}}^j = p_{\bar{B}_s^0}^i \, \delta_{ij} \, \epsilon_{D_{s1}}^j = p_{\bar{B}_s^0}^i \, \epsilon_{D_{s1}}^i = - p_{\bar{B}_s^0}^\mu \epsilon_{\mu,D_{s1}},
\end{eqnarray}
where in the last step we have written the product in a covariant way, which is valid in any frame of reference.

When we  evaluate $|t_{\bar{B}_s^0 \to K^- D_{s1}}|^2$ and sum over polarizations of the $D_{s1}$ we will have
\begin{eqnarray}
    \sum_{\mathrm{pol}} p_{\bar{B}_s^0}^\mu \, \epsilon_{\mu,D_{s1}} \, \epsilon_{\nu, D_{s1}} \, p_{\bar{B}_s^0}^\nu &=& \left( -g_{\mu\nu} + \frac{p_{D_{s1},\mu} \, p_{D_{s1},\nu}}{M^2_{D_{s1}}} \right) \, p_{\bar{B}_s^0}^{\mu} \, p_{\bar{B}_s^0}^{\nu} \nonumber \\ [2mm]
    &=& -M_{\bar{B}_s^0}^2 + \frac{1}{M_{D_{s1}}^2} \, \left( p_{D_{s1}} \cdot p_{\bar{B}_s^0} \right)^2,
\end{eqnarray}
which in the $\bar{B}_s^0$ rest frame is
\begin{equation}
    -M_{\bar{B}_s^0}^2 + \frac{1}{M_{D_{s1}}^2} \left( M_{\bar{B}_s^0} \cdot E_{D_{s1}} \right)^2 = \frac{M_{\bar{B}_s^0}^2}{M_{D_{s1}}^2} \cdot \vec{p}_{D_{s1}}^{\;2} = \frac{M_{\bar{B}_s^0}^2}{M_{D_{s1}}^2} \cdot \vec{p}_{K^-}^{\;2}
\end{equation}
where
\begin{equation}
    \label{eq:pk}
    p_{K^-, \,i} = \frac{\lambda^{1/2}(M_{\bar{B}_s^0}^2, M_{K^-}^2, M_{D_{s1, \,i}}^2)}{2M_{\bar{B}_s^0}}, ~~~~
    [i = D_{s1}(2460), D_{s1}(2536)] 
\end{equation}
and the decay widths are given by 
\begin{eqnarray}
    \Gamma_{i} = \frac{1}{8\pi} \, \frac{1}{M_{\bar{B}_s^0}^2} \, p_{K^-, \,i} \, \sum_{\mathrm{pol}} |t_{D_{s1}}^{(i)}|^2.
\end{eqnarray}
with $p_{K^-,\,i}$ given by Eq.~\eqref{eq:pk} for each $D_{s1}$ state, and 
\begin{eqnarray}
    \sum_{\mathrm{pol}}|t_{D_{s1}}^{(1)}|^2 &=& \mathcal{C}^2 \, \frac{M_{\bar{B}_s^0}^2}{M_{D_{s1,1}}^2} \; \vec{p}_{K^-,1}^{\; 2} \cdot \left|G_{D^* K}(\sqrt{s}=2460)\cdot g_{D_{s1},D^*K}\right|^2, \\[2mm]
    \sum_{\mathrm{pol}}|t_{D_{s1}}^{(2)}|^2 &=& \mathcal{C}^2 \, \frac{M_{\bar{B}_s^0}^2}{M_{D_{s1,2}}^2} \; \vec{p}_{K^-,2}^{\; 2} \cdot \left|G_{DK^*}(\sqrt{s}=2536) \cdot g_{D_{s1}^\prime,DK^*}\right|^2,
\end{eqnarray}
with $\mathcal{C}$ an arbitrary constant containing $\mathcal{C}^\prime$ in Eqs.~\eqref{eq:t2460} and \eqref{eq:t2536},  
that will cancel in the ratio for $D_{s1}(2536)$ and $D_{s1}(2460)$ production.

\subsection{Determination of the couplings and $G$ functions}

Here we divert a bit from the procedure followed in Ref.~\cite{Gamermann:2007fi} and use the extension of the local hidden gauge approach, exchanging vector mesons (implicit in Ref.~\cite{Gamermann:2007fi}) and regularizing the loops with the cutoff method, which is recommended when dealing with $D$ mesons \cite{Wu:2010rv}.
Also, unlike in Ref.~\cite{Gamermann:2007fi}, where a global perspective of all axial vector mesons was presented by means of a common subtraction constant in dimensional regularization, here we can tune the cutoff in the cutoff method to get the precise mass of the $D_{s1}$ states.

With the channels $D^{*0} K^+$, $D^{*+} K^0$ or $D^0K^{*+}$, $D^+ K^{*0}$ we get the same potential as obtained in Ref.~\cite{Ikeno:2023ojl} for the channels $D^0 K^+ (1)$, $D^+ K^0 (2)$, which taking $M_V$ as, approximately, the average of the $\rho$, $\omega$ and $\phi$ masses is given by
\begin{eqnarray}
    \label{eq:Vij}
    && V_{ij} = C_{ij} \,g^{\prime \, 2}\, (p_1 + p_3) \cdot (p_2 + p_4); \\[2mm]
    && g^\prime=\frac{M_V}{2f},~~ M_V=800\mev, ~~f=93\mev, \nonumber
\end{eqnarray}
with $f$ the pion decay constant, and
\begin{eqnarray}
    \label{eq:cij}
    C_{ij} = \left(
    \begin{array}{cc}
    -\frac{1}{M_V^2} & -\frac{1}{M_V^2}\\[2mm]
    -\frac{1}{M_V^2} & -\frac{1}{M_V^2}
    \end{array}
    \right),
\end{eqnarray}
which projected over $S$-wave gives
\begin{eqnarray}
    (p_1 + p_3) \cdot (p_2 + p_4) = \frac{1}{2} \left[3s - (M^2 + m^2 + M^{\prime \, 2} + m^{\prime \, 2}) - \frac{1}{s}(M^2 - m^2)(M^{\prime \, 2} - m^{\prime \, 2}) \right],
\end{eqnarray}
with $M$, $M^\prime$ the initial, final vector masses, and $m$, $m^\prime$ the initial, final pseudoscalar masses.
Taking average masses between the $D$, $K$ ($D^*, K^*$) multiplets we can go to the isospin basis using the wave function of Eqs.~\eqref{eq:IKDs} and \eqref{eq:IDKs} and get the combination $(V_{11} + V_{22} + 2V_{12})/2 = 2V_{11}$ for the $I=0$ states, hence
\begin{equation}
    \label{eq:V0}
    V = -\frac{2}{M_V^2} \,(p_1 + p_3) \cdot (p_2 + p_4) g^{\prime \, 2},
\end{equation}
and the scattering matrix will be given by
\begin{equation}
    \label{eq:bs}
    T = \frac{V}{1-VG} = \frac{1}{V^{-1} - G},
\end{equation}
with
\begin{equation}
    \label{eq:G}
    G(s) = \int_{|\vec{q}\,| < q_{\rm max}} \frac{ \mathrm{d}^3 q}{(2\pi)^3} \;\frac{\omega_1+\omega_2}{2\omega_1\omega_2} \;\frac{1}{s-(\omega_1+\omega_2)^2 + i\epsilon},
\end{equation}
with $\omega_i=\sqrt{m^2_i + \vec{q}^{\;2}}$.

We should recall that we have eliminated some coupled channels of Table~\ref{tab:axial}.
This can be done, as shown in Refs.~\cite{Aceti:2014ala,Hyodo:2013nka,Wang:2022pin}, with the consequence that the elimination of some channels reverts into an additional attraction in the remaining ones.
Indeed, in the case of two channels, $1$ and $2$, if we eliminate the less important channel $2$, we can get the same $T_{11}$ scattering amplitude using an effective potential \cite{Wang:2022pin}
\begin{equation*}
    V_{11, \,{\rm eff}}=V_{11} + \dfrac{V^2_{12}\, G_2}{1-V_{22}\, G_2} \simeq V_{11} +V^2_{12}\, G_2,
\end{equation*}
and since $G_2$ is negative, the effective potential ($V_{11}<0$) gets enhanced.
Then it should not be surprising that we need to increase a bit the potential of Eq.~\eqref{eq:V0} to get a proper binding. Alternatively we can also increase $q_{\rm max}$ in Eq.~\eqref{eq:G}, to get extra binding.
Indeed, increassing $q_{\rm max}$ makes the negative strength of $G$ bigger, and, hence, this is equivalent to making the strength of $V^{-1}$ (which is negative) smaller, which means increasing the strength of $V$.

\section{Results}
\label{sec:result}
With the former discussion, we present results with two scenarios:
\begin{enumerate}
    \item Take the potential of Eq.~\eqref{eq:V0} and fine tune $q_{\rm max}$ to get the pole at the right energy. We need $q_{\rm max} = 1025 \mev$ for the $DK^*$ interaction and $q_{\rm max} = 820 \mev$ for the $D^* K$ interaction to get the poles at the physical masses. The couplings obtained from the $T$ matrix at the poles, as $g_i^2 / (s-s_0)$, with $s_0$ the square of the resonance mass, are calculated as
    \begin{eqnarray}
    g_i^2 = \lim_{s \to s_0} (s-s_0) T_{ii} = \frac{1}{\frac{\partial}{\partial s} (V^{-1} - G_i)},
    \end{eqnarray}
    where L'H$\hat{\rm o}$pital's rule has been used in the second step of the equation.
    We obtain the values
    \begin{eqnarray}
    \label{eq:gc1}
    |g_{D_{s1}(2460),K D^*}| &=& 12107 \mev, \\[2mm]
    \label{eq:gc2}
    |g_{D_{s1}(2536),D K^*}| &=& 20234 \mev, 
    \end{eqnarray}
    and then the ratio $R$, defined as
    \begin{equation}
    \label{eq:R}
    R=\frac{\Gamma(\bar{B}_s^0 \to K^- D_{s1}(2460)^+)}{\Gamma(\bar{B}_s^0 \to K^- D_{s1}(2536)^+)},
    \end{equation}
    has the value
    \begin{equation}
        R=0.7.
    \end{equation}
    \item Here we take values of $q_{\rm max}$ in the line as used in other works with $q_{\rm max}$ from $750 \mev$ to $800 \mev$ and multiply the potential by a factor to get the poles at the right place. We multiply $V$ by $\alpha$ for the $DK^*$ interaction and by $\beta$ for the $D^* K$ interaction. The results can be seen in Table~\ref{tab:R}.
    \begin{table}[htb]
    \centering
    \renewcommand{\arraystretch}{1.2}
    \caption{The ratios of $R = \Gamma(\bar{B}_s^0 \to D_{s1}(2460)^+ K^-) / \Gamma(\bar{B}_s^0 \to D_{s1}(2536)^+ K^-)$ with different assumptions. The $q_{\rm max}$ and couplings $g$ are in units of MeV.}
    \begin{tabular}{c|c|c|c|c|c}
    \hline\hline
    ~~~~$q_{\rm max}$~~~~ & ~~~~$\alpha$~~~~ & ~~~~$\beta$~~~~ & \,$|g_{D_{s1}(2536),DK^*}|$\, & \,$|g_{D_{s1}(2460),KD^*}|$\, & ~~~~$R$~~~~ 
    \\ \hline\hline
    750 & 1.717 & 1.113 & 24863 & 12448 & 1.17 
    \\ \hline
    770 & 1.635 & 1.078 & 24393 & 12346 & 1.16 
    \\ \hline
    800 & 1.524 & 1.030 & 23737 & 12203 & 1.14 
    \\ \hline\hline
    \end{tabular}
    \label{tab:R}
    \end{table}
    We can see that we get different values than before. 
\end{enumerate}
We get results with two scenarios
\begin{equation}
    R ({\rm fixed ~potential}) =0.7,~~~~~
    R ({\rm fixed~} q_{\rm max}) =1.16.
\end{equation}
There is no compelling reason to prefer one result over the other and thus we should accept these two results as indicative of the uncertainties of our approach.
The recent measurement of LHCb on the $\bar{B}_s^0 \to K^+ D_{s1}(2536)^-$ reaction indicates that the same work for the $D_{s1}(2460)$ is also possible.
The future measurement of this decay channel would provide a valuable test for the nature of two $D_{s1}(2460)$ and $D_{s1}(2536)$ resonances.

\section{Conclusions}
\label{sec:concl}
The recent experimental measurement of the $\bar{B}_s^0$ decay to $K^\pm D_{s1}(2536)^\mp$ by the LHCb collaboration \cite{LHCb:2023eig} prompted the idea of the present work, evaluating the decay rates for the related $D_{s1}(2460)$ resonance. 
The two resonances are described from the molecular picture from the interaction of coupled channels, where the most important are the $D^*K$ for the $D_{s1}(2460)$ resonance and the $DK^*$ for the $D_{s1}(2536)$ resonance. 
We are able to determine the ratio of the branching ratios for $\bar{B}_s^0 \to K^- D_{s1}(2460)^+$ and $\bar{B}_s^0 \to K^- D_{s1}(2536)^+$ for which we get a value of the order of unity.  
The measurement of the decays for the $D_{s1}(2536)$ production indicates that the measurement of the same decay leading to the $D_{s1}(2460)$ resonance should also be possible. With the results of the present paper, this measurement should provide an important test for the molecular picture of these two resonances and the present work should provide an incentive for this experimental work to be performed.

\begin{acknowledgments}
    This work is partly supported by the National Natural Science Foundation of China under Grant No. 11975083, No. 12365019 and No. 12075019, the Central Government Guidance Funds for Local Scientific and Technological Development, China (No. Guike ZY22096024), and the Jiangsu Provincial Double-Innovation Program under Grant No.~JSSCRC2021488, and the Fundamental Research Funds for the Central Universities. 
    This work is also partly supported by the Spanish Ministerio de Economia y Competitividad (MINECO) and European FEDER
    funds under Contracts No. FIS2017-84038-C2-1-P B, PID2020-112777GB-I00, and by Generalitat Valenciana under contract
    PROMETEO/2020/023.
    This project has received funding from the European Union Horizon 2020 research and innovation
    programme under the program H2020-INFRAIA-2018-1, grant agreement No. 824093 of the STRONG-2020 project.
\end{acknowledgments}




\end{document}